\documentclass{aa}
\usepackage{graphicx}
\usepackage{txfonts}
\usepackage{epsfig}
\def\ltsima{$\; \buildrel < \over \sim \;$}
\def\gtsima{$\; \buildrel > \over \sim \;$}
\def\simlt{\lower.5ex\hbox{\ltsima}}
\def\simgt{\lower.5ex\hbox{\gtsima}}

\begin{document}

   \title{$Suzaku$ observation of the Phoenix Galaxy}

   \author{G. Matt \inst{1},  S. Bianchi\inst{1}, 
H. Awaki\inst{2}, A. Comastri\inst{3},
M. Guainazzi\inst{4}, K. Iwasawa\inst{3}, E. Jimenez-Bailon\inst{5}, F. Nicastro\inst{6}}

   \offprints{G. Matt, \email{matt@fis.uniroma3.it} }

   \institute {$^1$Dipartimento di Fisica, Universit\`a degli Studi Roma Tre, 
via della Vasca Navale 84, I-00146 Roma, Italy \\
$^2$ Department of Physics, Faculty of Science, Ehime University, Bunkyo-cho, Matsuyama, Ehime 790-8577, Japan\\
$^3$ INAF - Osservatorio Astronomico di Bologna, via Ranzani 1, 40127 Bologna, Italy \\
$^4$ European Space Astronomy Center of ESA, Apartado 50727, 28080 Madrid, Spain \\
$^5$ Instituto de Astronomía, UNAM, Apartado 70264, 04510 Ciudad de México, Mexico\\
$^6$ INAF - Osservatorio Astronomico di Roma, via Frascati 33, 00040 Monteporzio Catone, Roma, Italy 
}

   \date{Received / Accepted }

   \abstract{In recent years, several Seyfert 2 galaxies have been discovered 
that change state when observed in X-rays a few years apart, switching from Compton-thin to 
reflection-dominated or viceversa.}
{We observed a member of this class of ``Changing-look'' sources, the Phoenix Galaxy, with $Suzaku$, with the
aim of better understanding the nature of the variations.}
{The $Suzaku$ spectrum was analyzed, and the results compared with previous ASCA and XMM-$Newton$ observations. }
{The source was caught in a Compton-thin state, as in XMM-$Newton$,  but differently from ASCA. Comparing the $Suzaku$
and XMM-$Newton$ observations, a variation in the column density of the absorber on a time scale of years
is discovered. A similar change, but on much shorter time scales (i.e. ks)  
may also explain the count-rate variations during the $Suzaku$ observations. 

A soft excess is also present, likely due to continuum and line emission from photoionized circumnuclear matter.
 }{}

\authorrunning{G. Matt et al. }
\titlerunning{$Suzaku$ observation of the Phoenix Galaxy}

   \maketitle
%
%________________________________________________________________

\section{Introduction}

In recent years, it has been discovered that several Seyfert 2 galaxies 
change state when observed in X-rays a few years apart, switching back and forth 
from Compton-thin (i.e. absorbed
by line-of-sight matter with column densities lower than 
$\sigma_{T}^{-1}$=1.5$\times10^{24}$ cm$^{-2}$)
to reflection--dominated (Guainazzi 2001; Guainazzi et al. 2002; Matt, Guainazzi \& Maiolino 2003).
There are at present 6 objects in this 
class, i.e. about 10\% of the total population of reflection-dominated 
AGN in the local Universe (Guainazzi et al. 2005). 

The Compton reflection dominated
spectrum is usually assumed to be the result of Compton-thick absorption, the reflection being due
to the far side of the obscuring torus (e.g. Matt 2002 and references therein). Therefore,
the most obvious explanation for the change of state 
is the variation in the column density of the absorber.
This is indeed the case in one changing-look source, NGC~1365 (Risaliti et al. 2005), where
the column density is varying on very short time scale. However,
in another case, NGC~2992, this behaviour is clearly due to huge flux variations in the
nuclear radiation (Gilli et al. 2000): when the flux is very low, the reflection component (in this
case clearly arising from rather distant matter) remains the only visible component, an echo of the
source past activity. (This is what happened also in another famous source, NGC~4051, Guainazzi et al.
1998, which differs only because in the high flux state it is a Seyfert 1 instead of a Compton-thin
Seyfert 2, and on a much lower flux scale also to our own Galactic centre, Koyama et al. 2008 
and references therein). 
It is worth  noting that in this scenario the Compton-thick reflecting matter and the
Compton-thin absorber are likely to be associated with different regions, the latter possibly 
being related to the host galaxy (e.g. Lamastra et al. 2006).

In the other sources of this class, the nature of the variations is still unclear, due to
the lack of a proper monitoring campaign and/or of hard X-ray measurements.  Here we report
on the $Suzaku$ observation of one of the  brightest ``Changing-look'' source, the ``Phoenix Galaxy''.

The Phoenix Galaxy (as christened by Guainazzi et al., 2002, now an NED recognised name; 
a.k.a. UGC~4203 and Mkn~1210) is a Seyfert 2 at  $z$=0.0135.
It was observed by XMM--Newton on May 2001 (Guainazzi et al. 2002), unveiling
an X--ray bright nucleus (observed 2-10 keV flux of about 10$^{-11}$ erg cm$^{-2}$ s$^{-1}$),
absorbed by cold matter with  $N_H \simeq 2 \times 10^{23}$~cm$^{-2}$. 

In an ASCA observation performed about five and half years
earlier (Awaki et al. 2000, Guainazzi et al. 2002), the prominent iron line ($EW \simeq 1$~keV) and
the factor--of--5 lower 2--10~keV flux indicated instead 
a reflection--dominated spectrum, with the nuclear emission too faint to be visible.
The limited bandpass of ASCA could not permit checking whether the disappearing of the nucleus
below 10 keV was due to an increase of the column density into a moderately Compton-thick regime.
The lower limit to a putative absorber is  about 10$^{24}$ cm$^{-2}$.

\section{Observation and data reduction}

The Phoenix Galaxy was observed by $Suzaku$ on 2007  May 2. X-ray Imaging Spectrometer (XIS) 
and Hard X-ray Detector (HXD) event files were reprocessed with the latest calibration files 
available (2008-07-09 release), using \textsc{ftools} 6.5 and Suzaku software Version 9 and
adopting standard filtering procedures. Source and background spectra for all the three XIS 
detectors were extracted from circular regions of 2.9 arcmin radius, avoiding the calibration 
sources. Response matrices and ancillary response files were generated using \textsc{xisrmfgen} 
and \textsc{xissimarfgen}. We downloaded the ``tuned'' non-X-ray background (NXB) for our HXD/PIN 
data provided by the HXD team and extracted source and background spectra using the same good 
time intervals. The PIN spectrum was then corrected for dead time, and the exposure time of the 
background spectrum was increased by a factor 10, as required. Finally, the contribution
from the cosmic X-ray background (CXB) was subtracted from the source spectrum, simulating it 
as suggested by the HXD team. The final net exposure times are 59 ksec for the three XIS spectra 
and 46 ksec for the HXD/PIN (the source was not detected with the HXD/GSO). 

We also rereduced the XMM-$Newton$ data again (\textsc{obsid} 0002940701 - 5 May, 2001) with the 
latest calibration files. The observation was performed with the EPIC (European Photon Imaging 
Camera) CCD cameras, the p-n, and the two MOS (metal oxide semiconductors), operated in full window 
and thin filter. Data were reduced with SAS 7.1.0, and screening for intervals of flaring particle 
background was done consistently with the choice of extraction radii, in an iterative process 
based on the procedure to maximise the signal-to-noise ratio described by Piconcelli et al. 
(2004). The p-n was heavily affected by high background and resulted in a very short net 
exposure time 
(see G02) so will not be used in this paper. On the other hand, the net exposure time for the 
MOS cameras is 7634 s, adopting a source extraction radius of 14 arcsec and including patterns 
0 to 12. The background spectra were extracted from source-free circular regions with a radius 
of 50 arcsec. Since the two MOS cameras were operated with the same mode, we co-added MOS1 and 
MOS2 spectra, after having verified that they agree with each other and with the summed spectrum.

\section{Data analysis}

\subsection{Imaging and timing analyses}

\begin{figure}
\epsfig{file=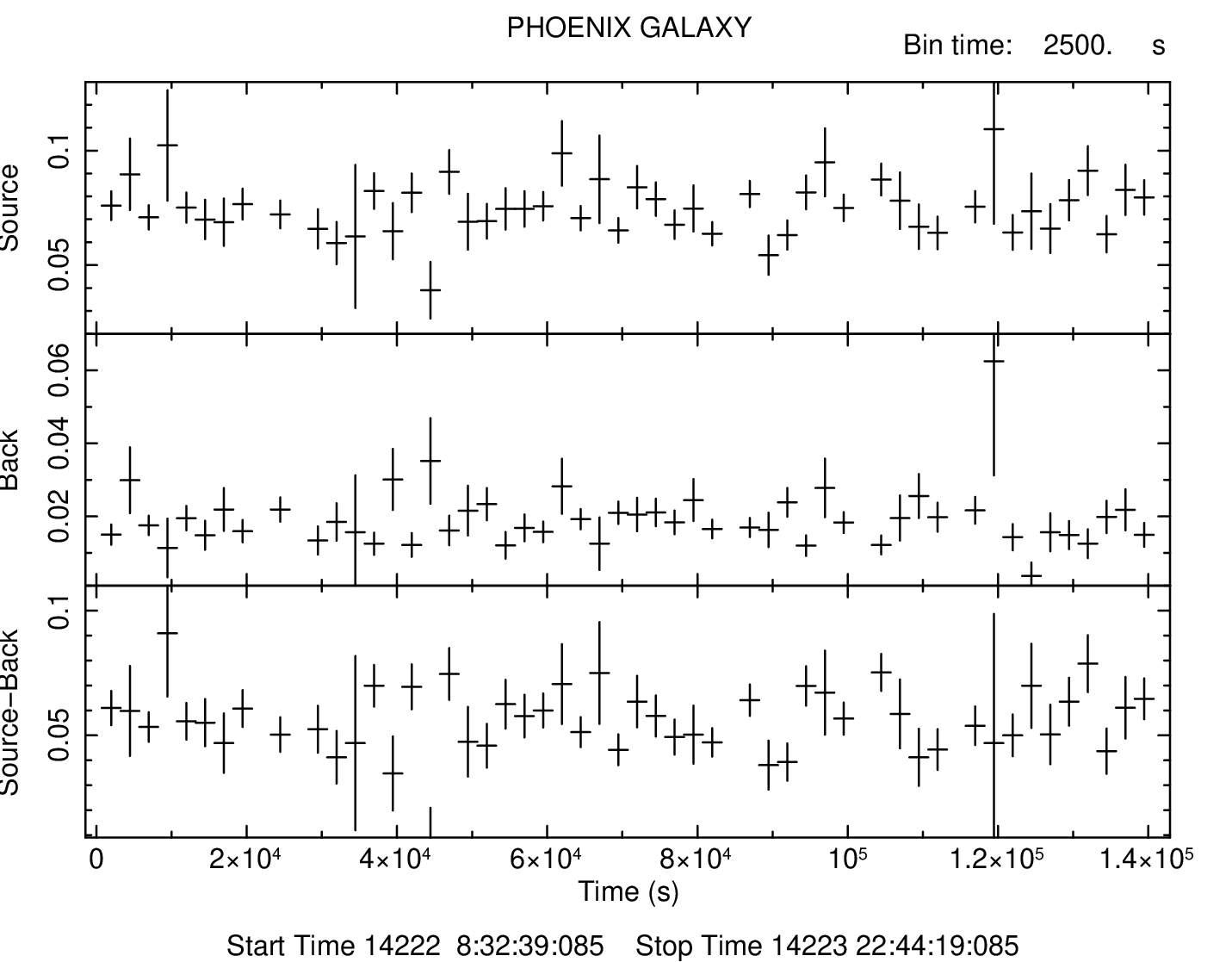,width=6cm,angle=-90}
\caption{The 0.3-2 keV total (upper panel), background (middle panel), and 
net source (lower panel)  XIS0+XIS1+XIS3 light curves for the$Suzaku$ observations.}
\label{lc1}
\end{figure}

\begin{figure}
\epsfig{file=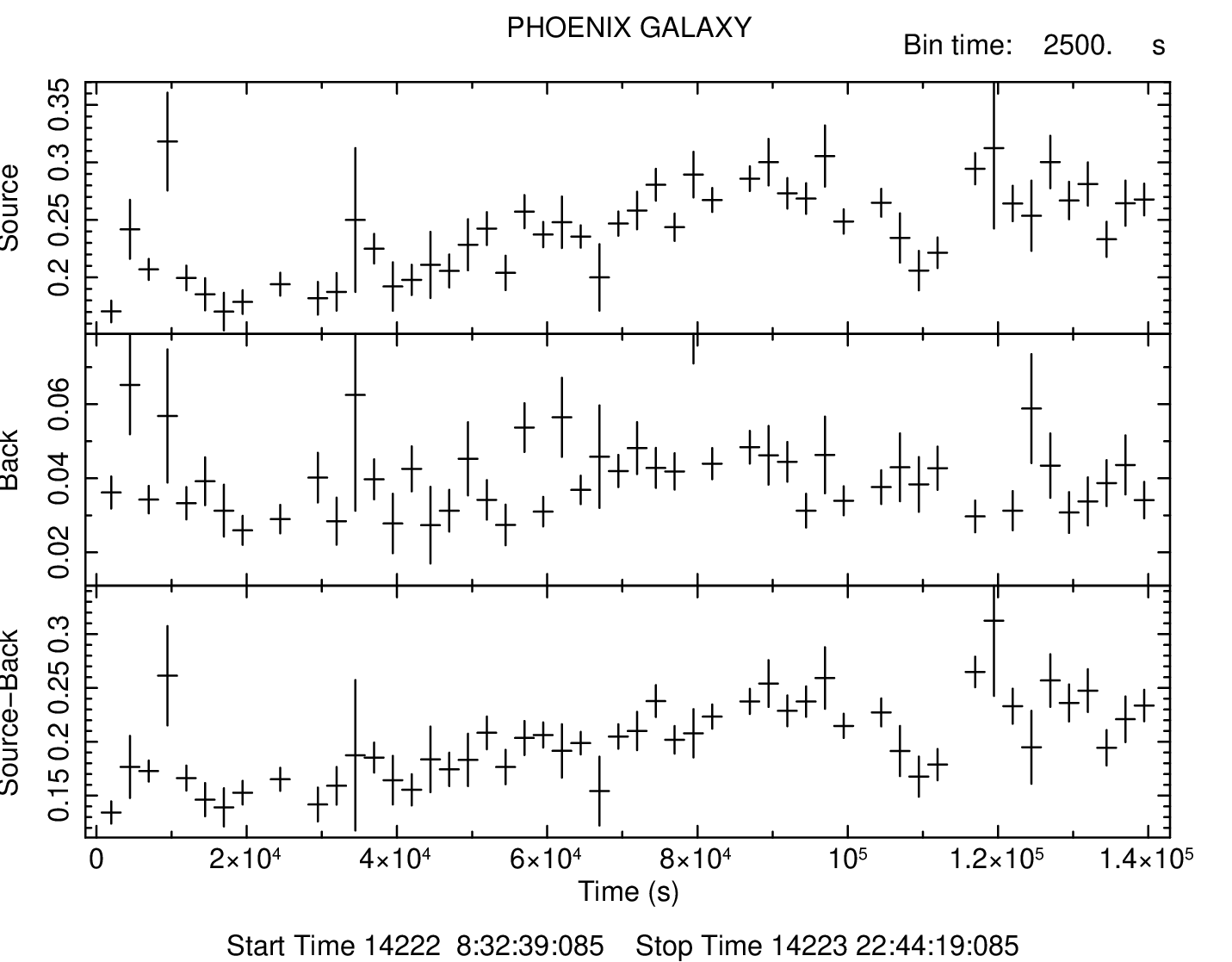,width=6cm,angle=-90}
\caption{The same as in the previous figure, but for the 2-10 keV band.}
\label{lc2}
\end{figure}

The only source present in the field of view of the XIS were the target itself. No bright
sources are found in the XMM--$Newton$ images, either. 
We extracted light curves from the XIS and PIN instruments. The 0.3-2 keV
and 2-10 keV  XIS0+XIS1+XIS3 light curves are shown in Figs.~\ref{lc1} and \ref{lc2}. While the
0.3-2 keV light curves do not show any significant variation in the source emission,
a clear increase in the flux is observed in the 2-10 keV light curve. This suggests
that, above a few keV, the nucleus is directly visible, therefore that the source
has been caught in a Compton-thin state. 

No significant variability is observed in the PIN. The source is detected up to about 50 keV,
with a 13-50 keV count rate of 0.074$\pm$0.003 counts/s.

\subsection{Spectral analysis. Suzaku}

For the spectral analysis, we used the full complement of $Suzaku$ instruments
apart from the XIS2, no longer in use at the time of the observation, and the 
GSO, as the source is not detected with this instrument. 
We added a normalization constant
for each instrument, fixing the value for XIS0 to 1, and to 1.18
for the PIN, as appropriate for data taken at the HXD nominal position. 
The values for the XIS1 and XIS3 instruments were left
free to vary, and turned out to be about 1.02 and 0.97, respectively.
The energy bands are 0.5-10 keV for the XIS and 13-50 keV for the HXD/PIN. 

The fits were performed using the {\sc xspec 12} software package. Errors refer
to 90\% confidence level for one interesting parameter.

\subsubsection{The source status}

Following previous results (G02), and with the aim of characterising
the status of the source, we started fitting the data above
4 keV for simplicity in order to minimise the contribution from the soft X-ray component, 
whatever its origin (see below). We adopted a baseline model composed of:
a power law; a Compton reflection (CR) component, 
with metal abundances fixed to the cosmic value (Anders \& Grevesse 1989)
and the inclination angle fixed to 30$^{\circ}$; a neutral and narrow iron line.
All these components are seen through an absorber at the
redshift of the source
(absorption model {\sc zphabs}\footnote{Abundances set to the
Anders \& Grevesse 1989 values, cross sections from  Balucinska-Church \& McCammon (1992). See
{\tt http://heasarc.gsfc.nasa.gov/docs/xanadu/xspec/ } for details.}; 
Thomson scattering is also considered including the model {\sc cabs}). 
The fit is perfectly acceptable, $\chi^2_r/d.o.f.$=0.92/150. The best fit results are summarised
in Table~\ref{fit}. 

An equally good fit ($\chi^2_r/d.o.f.$=0.91/150) is obtained if the reflection component is left outside the absorber.
In this case the $N_H$ is a bit larger (4.19$\times$10$^{23}$ cm$^{-2}$), the power law index is very similar
(1.84), and the reflection component smaller (1.26).

\begin{table*}
\begin{center}
\caption{Best fit parameters for the baseline model, for $E>$4 keV. 
}
\begin{tabular}{ccccccccc}
\hline
& & & & & & \\ 
 & $\Gamma$ & $N_H$  &  $R$ & E$_{line}$ & F$_{line}$ & EW & $\chi^2_r$/d.o.f. & Flux (XIS0; 4-10 keV)  \\
 &  & (10$^{23}$ cm$^{-2}$) & & (keV) & 10$^{-5}$ ph cm$^{-2}$ s$^{-1}$ & (eV) &  
& 10$^{-12}$  erg cm$^{-2}$ s$^{-1}$ \\
& & & & & & & & \\ 
\hline
& & & & & & & & \\ 
Total & 1.87$\pm0.18$  &  3.30$^{+0.22}_{-0.23}$ & 2.0$^{+1.7}_{-0.9}$ & 
6.398$\pm0.009$ & 4.34$^{+0.55}_{-0.53}$ & 206$\pm$26 & 0.91/150 & 4.26 \\
& & & &  & & & & \\ 
Low & 1.90$\pm0.33$  &  3.23$^{+0.51}_{-0.66}$ & 3.0$^{+9.8}_{-2.1}$ & 
6.389$^{+0.018}_{-0.024}$ & 3.90$^{+0.91}_{-0.86}$ &  228$\pm$53 & 1.00/62 & 3.53 \\
& & & &  & & & & \\ 
High & 1.87$\pm0.22$  &  3.26$^{+0.29}_{-0.28}$ & 1.6$^{+1.9}_{-0.9}$ & 
6.404$^{+0.010}_{-0.015}$ & 4.65$^{+0.76}_{-0.72}$ & 200$\pm$33 & 0.99/111 & 4.77 \\
& & & &  & & & & \\ 
\hline
\end{tabular}
\noindent
\label{fit}
\end{center}
Note: from left to right: the index of the
power-law ($\Gamma$); the column density of the intrinsic absorber; the relative normalization
of the Compton reflection component, $R$; the centroid energy, E$_{line}$, flux, F$_{line}$, and
equivalent width (EW) of the iron line; the reduced $\chi^2$ and the d.o.f.
\end{table*}

As already suggested by the timing analysis,  
the source is Compton-thin, as in the XMM--$Newton$ observation (Guainazzi et al. 2002)
and differently from the ASCA observation, when it was Compton-thick. In fact, when fitting the
$Suzaku$ spectrum with a pure CR ($\Gamma$ fixed to 2), 
the fit is much worse, $\chi^2_r$/d.o.f.=1.80/152; moreover, the 
iron line EW, about 240 eV, is far too low, see e.g. Ghisellini et al. (1994); Matt et al. (1996).     

\begin{figure}
\epsfig{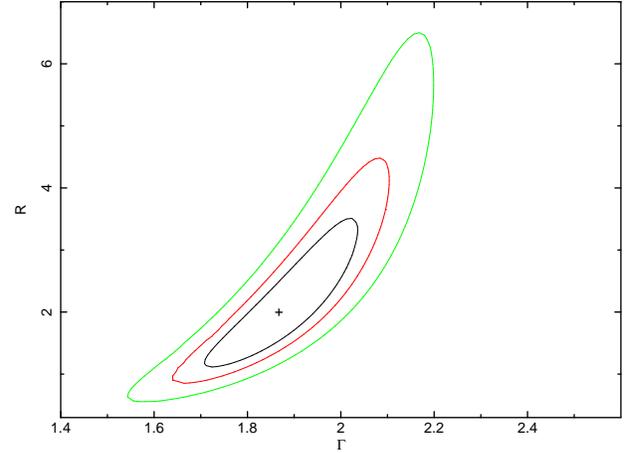}
\caption{The $\Gamma$-$R$ contour plot, fitting the $Suzaku$ spectrum above 4 keV.}
\label{contourplot1}
\end{figure}

The 4-10 keV flux is 4.3$\times$10$^{-12}$ erg cm$^{-2}$ s$^{-1}$ (XIS0), corresponding to a 2-10
keV unabsorbed flux of  1.9$\times$10$^{-11}$ erg cm$^{-2}$ s$^{-1}$ and luminosity of 
7.7$\times$10$^{42}$ erg s$^{-1}$ ($H_0$=70 km/s/Mpc).
The index of the power law, $\Gamma$, and the relative normalization of the Compton reflection
component, $R$, turned out to be highly correlated with each other; see Fig.~\ref{contourplot1}.
At the 2$\sigma$ level, $\Gamma$ can range from about 1.65 up to about 2.1,
with $R$ ranging correspondingly from about 1 up to 4. The value of the equivalent width
of the iron line, about 200 eV, strongly favours the low $\Gamma$, low $R$ scenario. In fact,
for cosmic abundancies such a value of the iron line EW would correspond to a value of $R$ in
between 1 and 1.5 (e.g. Matt et al. 1991), not forgetting that part of the line (up to 50-60
eV for large covering factors, e.g. Matt et al. 2003) may originate in the absorber
itself. 

Adopting the model {\sc zwabs} instead of {\sc zphabs} for absorption, rather different
best-fit values are found, due to the different cross sections and abundancies adopted
(Morrison \& McCammon 1982; Anders \& Ebihara 1982).
The column density remains about the same (3.26$^{+0.28}_{-0.33}\times$10$^{23}$
cm$^{-2}$) (as well as the quality of the fit: $\chi^2_r$/d.o.f.=0.94/150);
but $\Gamma$ is now  2.13$^{+0.16}_{-0.17}$, while $R$ becomes 5.2$^{+4.8}_{-2.3}$, formally
consistent within the errors with the values found using {\sc zphabs}, but not allowing for the low $R$
solution required by the iron line EW. 

The presence of a Compton-thin absorber and a Compton-thick reflection indicates that
more than one circumnuclear region does exist, as often observed in Seyfert 2s. To check
this, we also fitted the data with a model developed by one of us (HA) and his collaborators 
(``Ehime'' model, here-in-after: Ikeda et al. 2009).
The model, based on MonteCarlo simulations,
 assumes that the absorption and reflection come from the same material, 
a torus with a half-opening 
angle which is one of the fitting parameters. The fit is still acceptable, $\chi^2_r/d.o.f.$=1.13/150, 
but significantly worse than that with the Compton-thick reflector.
The best-fit column density is
about 3.7$\times$10$^{23}$ cm$^{-2}$, with an half opening angle of about 30$^{\circ}$ (but very poorly constrained
to be within 10$^{\circ}$  and 50$^{\circ}$) and $\Gamma\sim$1.5.
Allowing the absorbing and reflecting material to be different, not surprisingly, a fit as good as 
the one in Table~\ref{fit} is
recovered; the column density of the torus (now responsible only for the reflection) is 
4.7$\times$10$^{24}$ cm$^{-2}$, with a better determined half opening angle of 25$^{+12}_{-4}$ degrees, 
and a slightly larger inclination angle.

\subsubsection{Variability}

As noted earlier, the source flux changes during the observation. In particular, it is consistently
higher in the second half of the observation. To search for spectral variability, we divided the
observation into two parts, the dividing time at about 50 ks (elapsed time) 
from the start of the observation. The resulting spectra (from now on denoted as ``low''
 and  ``high'') have net exposures of 24 and 35 ks (XIS), and 4-10 keV count rates (XIS0) of 0.0445$\pm$0.0015
and  0.0638$\pm$0.0015 cts/s, respectively (to be compared with a count rate in the total spectrum
of 0.0577$\pm$0.0011). The 13-50 keV count rates in the PIN are 0.070$\pm$0.005 and 0.077$\pm$0.004.

The fits with the baseline  model (E$>$4 keV) are reported in Table~\ref{fit}. All spectral parameters
are consistent with one another within the errors, indicating that the variation may be entirely due to a change in 
the normalization of the intrinsic flux. Indeed, forcing the two column densities to be the same and allowing
the power law and the iron line normalizations, as well as the $R$, to vary, leads to an acceptable 
fit ($\chi^2$/d.o.f.=1.14/69). In the fit only XIS0 and PIN
were used, to avoid complications in the fitting procedure due to the normalization constants between the three
XISs. The small (or no)  variability in the PIN is, in this fit, accounted for by the increase of 
$R$ in the low state (best-fit values of 2.2 and 1.3, even if the two values are consistent each other
within the very large errors),
as expected if the reflecting matter is at large distances from the black hole.

\begin{figure}
\epsfig{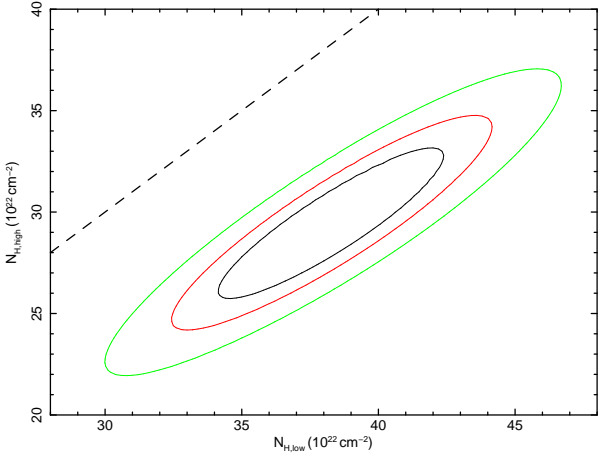}
\caption{Contour plot for the column densities in the ``low'' and ``high'' states,
assuming that the observed variability is entirely due to variations in the absorber (see text
for details). The dashed line refers to the equal $N_H$ relation.}
\label{contourplot2}
\end{figure}

On the other hand, the lack of  significant variability in the PIN
(the [13-50 keV]/[4-10 keV] hardness ratios in the two states are formally 
inconsistent with each other) may suggest that
the change is entirely due to column density variations. Indeed, fitting simultaneously the ``low'' and ``high'' spectra 
letting only the N$_H$  to vary, a better fit is found ($\chi^2$/d.o.f.=1.11/71).
The contour plot between the two column densities is shown in Fig.~\ref{contourplot2}, where it is seen
that in this scenario the column density varies by about 0.6$\times$10$^{23}$ cm$^{-2}$. 
Using all XIS detectors, a qualitatively similar result is found, but the error bars are 
significantly larger due to the above-mentioned cross-calibration uncertainties.

To summarise, the quality of the data is not good enough to firmly assess the nature
of the observed count rate variability, even if the $N_H$ variability seems to be slightly 
preferable, providing a better fit with only one parameter free to vary. 
In any case, because the spectral  variability is low, if any, 
for the sake of simplicity we only consider the total spectrum in the following.

\subsubsection{The iron line complex}

The iron line energy of 6.398$\pm0.009$ keV is perfectly consistent with neutral iron. If $\sigma$
is left free to vary, no significant improvement in the quality of the fit 
is found; an upper limit of about 60 eV can be put to $\sigma$. 
Similarly, no improvement is found by adding a Compton Shoulder (CS; see 
e.g. Sunyaev \& Churazov 1996; Matt 2002; Watanabe et al. 2003), 
assumed for simplicity to be a Gaussian with $\sigma$=40 eV centered at 6.32 eV (Matt 2002).

%The asymmetric broadening suggested by the possible presence of a CS
%may of course also be due to emission from a relativistic disc. We then substituted the
%Gaussian with the {\sc diskline} model, with the inclination angle fixed
% to 30$^{\circ}$, the outer radius
%to 10$^6$ gravitational radii and the emissivity law to 2.5. The fit is
%as good as the one with the CS
%($\chi^2_r$/d.o.f.=0.87/149), but the inner radius is basically  unconstrained, ranging from
%6 to 800 (best fit of 80) gravitational radii.

We also searched for the iron K$\beta$ (7.058 keV) and the nickel K$\alpha$ lines  (7.472 keV). 
The upper limit to the fluxes are 5.0$\times10^{-6}$ (i.e. 12\% of the iron K$\alpha$) 
and  5.8$\times10^{-6}$ (13\% of the iron K$\alpha$), respectively. The upper limit to the K$\beta$/K$\alpha$
is consistent with low ionization (e.g. Molendi et al. 2003), in agreement with the line centroid energy. 
The upper limit to the nickel K$\alpha$ line is not particularly constraining, being consistent with an Ni/Fe
ratio up to 3-4 times the cosmic value.

\begin{table}
\begin{center}
\caption{Fluxes of the main emission lines expected in a photoionized
plasma. }
\begin{tabular}{ccc}
\hline
 & & \\ 
 Line & Energy & Flux \\
  & (keV) & (10$^{-6}$ ph cm$^{-2}$ s$^{-1}$) \\
 & & \\ 
\hline
& & \\ 
O {\sc vii} K$\alpha$ & 0.57 & 29.1$^{+23.5}_{-23.7}$ \\
O {\sc viii} K$\alpha$ & 0.65 & 16.3$^{+10.6}_{-10.5}$ \\
O {\sc vii} RRC & 0.74 &  12.5$^{+4.6}_{-4.6}$ \\ 
O {\sc viii} RRC & 0.87 & 15.0$^{+4.4}_{-3.0}$ \\
Ne {\sc ix} K$\alpha$ & 0.91 & 4.9$^{+3.0}_{-4.5}$ \\
Fe {\sc xx} L & 0.965 & 9.6$^{+2.6}_{-7.1}$ \\
Ne {\sc x} K$\alpha$ & 1.02 & $<7.0$ \\
Fe {\sc xxii} L & 1.053 & 6.9$^{+2.0}_{-4.6}$ \\
Fe {\sc xxiii} L & 1.13 & 3.7$^{+2.5}_{-2.8}$ \\
Fe {\sc xxiv} L & 1.17 & $<3.3$ \\
Mg {\sc xi} K$\alpha$ & 1.34 & 1.5$^{+1.2}_{-1.2}$ \\
Si {\sc xiii} K$\alpha$ & 1.85 &  $<2.1$ \\
& & \\ 
\hline 
\end{tabular}
\label{lines}
\end{center}
\end{table}

\subsubsection{The soft excess}

\begin{figure}
\epsfig{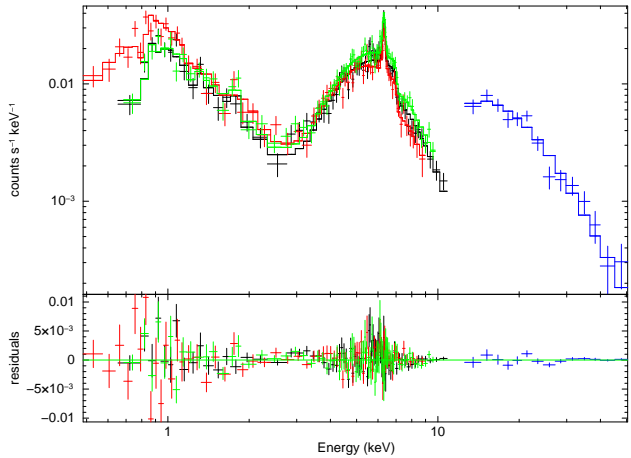}
\caption{Spectrum, best-fit model and resisuals for the $Suzaku$ observation. See text
for details.}
\label{spectrum}
\end{figure}

A significant emission in excess of the absorbed power law is clearly apparent in the spectrum
(see Fig.~\ref{spectrum}). 
We first parametrized the soft excess with a power law (absorbed by the Galactic column, 
$N_{H,Gal}$=3.45$\times$10$^{20}$ cm$^{-2}$) with the index forced to be
the same as that of the hard power law, as expected if the soft emission is due to Thomson reflection
of the nuclear radiation. The fit is unacceptable ($\chi^2_r$/d.o.f.=1.71/231), 
with clear excesses below 2 keV. Allowing the two indices to vary independently, the fit slightly improves
but remains unacceptable ($\chi^2_r$/d.o.f.=1.64/230). This
is not suprising, as recent high resolution grating observations of Seyfert 2s have shown
that most of the soft X-rays are due to emission lines (e.g. Guainazzi \& Bianchi 2007, and references
therein). Indeed, adding a thermal plasma component ({\sc mekal}) dramatically improves the fit 
($\chi^2_r$/d.o.f.=1.07/227). The temperature is 0.75$^{+0.04}_{-0.07}$ keV, but  
the metal abundancies are unrealistically low (0.09$^{+0.01}_{-0.03}$) and the soft power
law very flat ($\Gamma_s$=0.12$^{-0.14}_{+0.19}$). Indeed, the soft power law (of unknown origin)
is required, because without it the fit is much worse ($\chi^2_r$/d.o.f.=1.61/229).

\begin{figure}
\epsfig{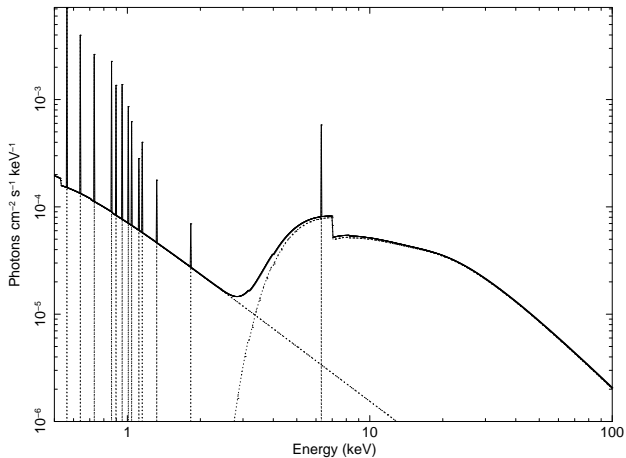}
\caption{The best-fit unfolded model. See text for details.}
\label{model}
\end{figure}

Alternatively, the lines may be emitted in photoionized, rather than collisionally ionized, plasma, 
as seems to be the case for most Seyfert 2s (Guainazzi \& Bianchi 2007). We therefore substituted the 
{\sc mekal} with many Gaussian, narrow-emission lines with the energy fixed to that of the most
important transitions (see Table~\ref{lines}), even if 
there is no direct evidence from eye inspections of the spectrum for most of them.
 The two power law indices were forced to be the same
(no significant improvement is found letting them vary independently).
The fit is good ($\chi^2_r$/d.o.f.=1.16/219), even if
not as good as the one with {\sc mekal}. 
The presence of O {\sc vii} and {\sc ovii} radiative recombination continua (RRC) is in agreement with
the assumption that the matter is indeed photoionized.  
The best fit values of $N_H$, $\Gamma$, $R$ and the iron line EW for this fit are
3.33$\times$10$^{23}$ cm$^{-2}$, 1.71, 1.28 and
218 eV. The spectrum and best fit models are shown in Fig.~\ref{spectrum} and  Fig.~\ref{model},
respectively, while the continuum parameters are summarized in Table~\ref{fitxmm}. From
Fig.~\ref{model} it can be seen that the hard and soft components have similar fluxes at about
3 keV.

The observed 0.5-2 keV flux is  2.7$\times$10$^{-13}$ erg cm$^{-2}$ s$^{-1}$, 3.3\% of the 
nuclear one, after correcting for absorption. This number is typical of Seyfert 2 galaxies
(Bianchi \& Guainazzi 2007). About 60\% of the soft flux is 
in the continuum, and the remaining 40\% in emission lines (adopting the 
photoionization scenario described above). Therefore, the column density of the photoionized
matter, $N_{H,scatt}$, is  (0.6$\times$0.033)$\sigma_T^{-1}(4\pi/\Delta\Omega)$, i.e.
 3$\times$10$^{22}(4\pi/\Delta\Omega)$ cm$^{-2}$, 
with $\Delta\Omega$ the solid angle subtended
by the reflecting matter to the ionizing continuum. In a biconical geometry, $\Delta\Omega$=4$\pi(1-cos\theta)$,
with $\theta$ the half-opening angle. If $\theta$=25$^{\circ}$, as suggested by the fit with the ``Ehime'' model,
$N_{H,scatt}\sim3\times$10$^{23}$  cm$^{-2}$.

\subsection{Spectral analysis. XMM-$Newton$}

The 0.5-10 keV XMM-$Newton$ spectrum was fitted with a model similar to the one described
in the previous paragraph, but letting the soft power-law index vary independently
of the hard one. The reason is that the quality of the XMM-$Newton$ spectrum is much lower
than that of $Suzaku$ due to the shorter exposure and the lack of p-n data, and the wealth
of lines likely present in the soft band cannot be fitted individually. Only a line at about 0.9 keV,
which is apparent in the data (see Guainazzi et al. 2002), is included. The best-fit values are summarised
in Table~\ref{fitxmm}.

The 4-10 keV flux is 9.5$\times$10$^{-12}$  erg cm$^{-2}$ s$^{-1}$, corresponding to a 2-10
keV unabsorbed flux of  2.6$\times$10$^{-11}$ erg cm$^{-2}$ s$^{-1}$ and luminosity of 
1.1$\times$10$^{43}$ erg s$^{-1}$.

\begin{table*}
\begin{center}
\caption{Best fit parameters for the broand band $Suzaku$ (photoionization scenario) and
XMM-$Newton$ spectra. 
}
\begin{tabular}{cccccccc}
\hline
& & & & & & & \\ 
& $\Gamma_s$ & $\Gamma_h$ & $N_H$  &  $R$ & E$line$ & EW & $\chi^2_r$/d.o.f. \\
& & & (10$^{23}$ cm$^{-2}$) & & (keV)   & (eV) &  \\
& & & & & & & \\ 
\hline
& & & & & & & \\ 
$Suzaku$ & 1.71$\pm0.12$ & (=$\Gamma_s$) & 3.33$^{+0.17}_{-0.15}$ & 1.28$^{+0.81}_{-0.56}$ & 6.397$\pm0.009$ &
 218$\pm$24 & 1.16/219 \\
& & & & & & & \\ 
\hline
& & & & & & & \\ 
XMM-$Newton$ & 2.49$\pm0.23$ & 1.62$^{+0.11}_{-0.23}$ & 1.76$^{+0.13}_{-0.21}$ & $<$2.1 &  6.44$\pm0.05$ 
& 133$\pm$63 & 1.21/70 \\
& & & & & & & \\ 
\hline
\end{tabular}
\noindent
\label{fitxmm}
\end{center}
Note: from left to right: the indices of the soft ($\Gamma_s$)
and hard ($\Gamma_h $) power laws; the column density of the intrinsic absorber; 
the relative normalization
of the Compton reflection component, $R$; the centroid energy, E$_{line}$ and
equivalent width (EW) of the iron line; the reduced $\chi^2$ and the d.o.f. 
\end{table*}

\section{Comparison between Suzaku and XMM-$Newton$ observations}

After comparing the $Suzaku$ and XMM-$Newton$ observations, we first note that 
the intrinsic column density is higher (about 
3.3$\times10^{23}$  against 1.8$\times10^{23}$ cm$^{-2}$). 
The difference is statistically significant 
(e.g., fixing
the $N_H$ in the XMM-$Newton$ ($Suzaku$) observation to be equal to the  $Suzaku$ (XMM-$Newton$)
best-fit value makes the fits much worse,  $\Delta\chi^2$=67 (38)). This suggests that the absorber is
relatively close to the nucleus, varying on timescales no longer than a few years.
 
The EW of the neutral iron line is about 200 eV in the $Suzaku$ observation, 
more than in  XMM--$Newton$ data (but still consistent within the errors), while the
line flux is about the same, i.e 4.5$\times10^{-5}$ ph cm$^{-2}$ s$^{-1}$ 
against 3.9$\times10^{-5}$ ph cm$^{-2}$ s$^{-1}$; the two values are consistent each other 
within the errors. (The same is true if the line is put {\it outside} the absorber, in the hypothesis
that the absorbing matter is closer to the black hole than the line emitting matter, in which case,
the line fluxes are 1.9$\times10^{-5}$ and 2.4$\times10^{-5}$ ph cm$^{-2}$ s$^{-1}$.)
The best-fit value of $R$ is also higher, but consistent within the errors. (It 
is very poorly constrained
in the XMM-$Newton$ data due to the limited bandwidth.)
The 2-10 keV XMM-$Newton$ unabsorbed flux is about 1.5 times larger than the $Suzaku$ one. 
As a result, we can conclude that the source was in a quite similar state during the $Suzaku$ and
XMM-$Newton$ observations, albeit somewhat fainter in the former.

\section{Discussion and conclusions}

The main results from the 2007 $Suzaku$ observation of the ``changing-look''
source, the Phoenix
Galaxy (the first observation in hard X-rays, as the source was not
observed by Beppo$SAX$), can be summarized as follows.

\itemize

\item The source was clearly in a Compton-thin state, as in  
2001, when it was observed by XMM--$Newton$ (Guainazzi et al. 2002), and differently than
in 1995, when $ASCA$  caught it in a  
reflection--dominated state (Awaki et al. 2000, Guainazzi et al. 2002).

\item Guainazzi et al. (2002) interpreted the change of state from 1995 to 2001 in terms of a switching--on 
of the nuclear emission, which was off in the ASCA observation (having ``re-birth'' from its
ashes, hence the ``Phoenix'' nickname). The (corrected for absorption) flux change from the XMM--$Newton$ to
the $Suzaku$ observations (when it was about a factor 1.5 fainter) may support this scenario. On 
the other hand, the line--of--sight column density is different in the two observations, $N_H$ being
almost a factor 2 higher in the $Suzaku$ data. The two values are inconsistent each other with high
statistical significance, so the hypothesis that the change of state was due to variations in
the absorber, as in NGC~1365 (Risaliti et al. 2005), cannot be ruled out, either.

\item Variations in the column density of the absorber on hourly time scales 
may also explain the count rate
variability observed during the $Suzaku$ observation. (This intepratation is somewhat 
supported by the discovery of column density
variations on weekly time scales in the $Chandra$ data of this source:
Risaliti et al., in preparation.)
However, an explanation in terms
of variations in the flux of the primary component cannot be excluded. 

\item There is a significant reflection component, which implies
a large amount of circumnuclear Compton--thick matter (e.g. Matt et al. 2003)
even when the source is Compton--thin. The fit is not able to tell us if the reflecting matter is
within or outside the absorber. Because the latter is varying on yearly time scales,
and possibly on much shorter ones, it could not be too far away from the nucleus. Therefore, 
if the reflector is within the absorber, the natural candidate  would
be the accretion disc,  the ``torus'' if it is outside.
Within the admittedly large (especially for the 
short XMM--$Newton$ observation) errors, the amount of the reflection component and of the iron
line are constant between 2001 and 2007, suggesting emission from distant matter. 

\item
The soft X-ray emission is fitted well by both a photoionised and a collisionally ionized 
material. However, the former solution is to be preferred, because it is more self-consistent
(the collisionally plasma model requires the presence of a (very flat) power law, too, and very low
abundancies) and because
of the detection of the O {\sc vii} and {\sc viii} radiative recombination continua.

\section*{Acknowledgements}

GM  and SB acknowledge financial support from ASI under grants I/023/05/0
and I/088/06/0.


\begin{thebibliography}{}

\bibitem[]{} Anders E., Ebihara M., 1982, Geo. Cosm. Acta, 46, 2363

\bibitem[]{} Anders E., Grevesse N., 1989, Geo. Cosm. Acta, 53, 197

\bibitem[]{} Arnaud K. A., 1996, in Jacoby G. H., Barnes J., eds, ASP Conf. Ser.Vol. 101, Astronomical Data Analysis Software and Systems V. Astron. Soc. Pac., San Francisco, p. 17

\bibitem[]{} Awaki H., Ueno S., Taniguchi Y, Weaver K.A., 2000, ApJ, 542, 175

\bibitem[]{} Baluinska-Church M., McCammon D., 1992, ApJ, 400, 699

\bibitem[]{} Bianchi S., Guainazzi M., 2007, AIP Conference Proceedings, Volume 924, pp. 822

\bibitem[]{} Gilli R., Maiolino R., Marconi A., et al., 2000, A\&A, 355, 485

\bibitem[]{} Ghisellini G., Haardt F., Matt G., 1994, MNRAS, 267, 743.

\bibitem[]{} Guainazzi M., Nicastro F., Fiore F., et al., 1998, MNRAS, 301, L1

\bibitem[]{} Guainazzi M., 2001, MNRAS, 329, L13

\bibitem[]{} Guinazzi M., Matt G., Fiore F., Perola G.C., 2002, A\&A, 388, 787

\bibitem[]{} Guainazzi M., Fabian A.C., Iwasawa K., Matt G., Fiore F., 2005a, MNRAS, 356, 295

\bibitem[]{} Guainazzi M., Matt G., Perola G.C.,  2005, A\&A, 444, 119

\bibitem[]{} Guainazzi M., Bianchi S., 2007, MNRAS, 374, 1290

\bibitem[]{} Ikeda S., Awaki H., Terashima Y., 2009, ApJ, in press

\bibitem[]{} Koyama K., Inui T., Matsumoto H., Tsuru T.G., 2008, PASJ, 60, 201

\bibitem[]{} Lamastra A., Perola G.C., Matt G., 2006, A\&A, 449, 551 

\bibitem[]{} Matt G., Perola G. C., Piro L., 1991, A\&A, 247, 25

\bibitem[]{}  Matt G., Brandt W.N, Fabian A.C., 1996, MNRAS, 280, 823

\bibitem[]{} Matt G., 2002, MNRAS, 337, 147

\bibitem[]{} Matt G., 2002, RSPTA, 360, 204

\bibitem[]{} Matt G., Guainazzi M., Maiolino R., 2003, MNRAS, 342, 422

\bibitem[]{} Molendi S., Bianchi S.,  Matt G., 2003, MNRAS, 343, L1

\bibitem[]{} Piconcelli E., Jimenez-Bailon E., Guainazzi M., et al., 2004, MNRAS, 351, 161

\bibitem[]{} Sunyaev R.A., Churazov E.M., 1996, Astr. Letters, 22, 648

\bibitem[]{} Watanabe S., Sako M., Ishida M., et al., 2003, ApJ, 597, L37




\end{thebibliography}
\end{document}